\documentclass[cameraready]{Interspeech}

\title{AnimeScore: A Preference-Based Dataset and Framework for Evaluating Anime-Like Speech Style}

\author[affiliation={1}]{Joonyong}{Park}
\author[affiliation={1}]{Jerry}{Li}

\address{
    $^1$ Spellbrush, USA
}

\email{jyjoon97@gmail.com, jerry@sizigistudios.com}

\keywords{Speech evaluation, Speech prediction model, Speech dataset, Corpus design, Voice likability}

\usepackage{comment}
\usepackage{multirow}
\usepackage{cite}
\usepackage{tabularx} 
\usepackage{graphicx}
\usepackage{subfig}
\newcommand{\myparagraph}[1]{\vspace{1mm}\noindent\textbf{#1}}

\begin{document}
\maketitle
\begin{abstract}
\vspace{-2mm}
Evaluating `anime-like' voices currently relies on costly subjective judgments, yet no standardized objective metric exists. A key challenge is that anime-likeness, unlike naturalness, lacks a shared absolute scale, making conventional Mean Opinion Score (MOS) protocols unreliable. To address this gap, we propose AnimeScore, a preference-based framework for automatic anime-likeness evaluation via pairwise ranking. We collect 15{,}000 pairwise judgments from 187 evaluators with free-form descriptions, and acoustic analysis reveals that perceived anime-likeness is driven by controlled resonance shaping, prosodic continuity, and deliberate articulation rather than simple heuristics such as high pitch. We show that handcrafted acoustic features reach a 69.3\% AUC ceiling, while SSL-based ranking models achieve up to 90.8\% AUC, providing a practical metric that can also serve as a reward signal for preference-based optimization of generative speech models.
\end{abstract}
\vspace{-2mm}
\section{Introduction}
\vspace{-1mm}
\emph{Anime} constitutes a major global industry where speech plays a central role in character expression.
Voice actors employ distinctive vocal techniques to portray animated characters, and the resulting notion of an ``anime-like'' voice is intuitively shared among creators and audiences.
Despite this shared intuition, no reproducible metric or automatic evaluation method for anime-likeness has been established, creating a practical bottleneck for developers of speech generation systems who must conduct costly listening tests to assess whether generated speech achieves the desired style, slowing iterative development.
The problem is further compounded by the nature of anime-likeness itself: unlike naturalness or intelligibility, it does not lend itself to absolute scoring on a shared numerical scale, as it is inherently multidimensional and lacks a perceptual anchor.

To address these challenges, we propose AnimeScore, a preference-based framework that jointly tackles large-scale data construction and automatic style prediction through pairwise ranking, focusing on Japanese speech.
Specifically, we (i)~collect 15{,}000 pairwise preference judgments from 187 evaluators to analyze the acoustic dimensions underlying their perceptual criteria and establish an interpretable baseline that quantifies the ceiling of handcrafted features, and (ii)~train ranking models on several SSL backbones, demonstrating that masked-prediction representations substantially surpass this baseline and closely reproduce human comparative judgments.
The resulting predictor directly enables two practical applications: a drop-in evaluation metric for rapid model screening without human raters, and a reward signal for reinforcement learning to optimize generative speech models toward a target style.
Our dataset metadata and implementation are publicly available~\footnote{github.com/sizigi/animescore}.

\vspace{-2mm}
\section{Related Work}
\subsection{Studies on Anime Voices}
\vspace{-2mm}
Acoustic properties of speech in anime have been examined in several prior studies.
Early work analyzed prosodic and acoustic differences across character roles~\cite{teshigawara2003}, and subsequent studies quantitatively characterized anime-specific voice quality using spectral and voice-quality descriptors~\cite{starr2015, utsugi2019}.
More recent work has further investigated whether there exist shared acoustic regularities across diverse styles~\cite{ishi2023}. Across these studies, tendencies have been reported including higher mean F0, wider pitch range, and shifts in voice quality related descriptors~\cite{starr2015, utsugi2019, ishi2023}. However, most of the findings are primarily descriptive: they do not directly yield a scalable scoring protocol or a trainable objective that can be used for comparison and iterative development of model under anime-style conditions.
\vspace{-3mm}
\subsection{Speech Evaluation and Automation}
\vspace{-1mm}
Subjective listening tests are the de facto standard for evaluating speech quality, but are expensive to run at scale and hard to repeat during the development of speech generation systems. 
This practical bottleneck has motivated extensive research on automatic prediction of human judgments, most prominently in the context of naturalness-oriented MOS regression.
Those MOS predictors commonly leverage self-supervised learning (SSL) speech representations to improve robustness and cross-domain generalization~\cite{wav2vec2, chen2022wavlm, pairwise_ranknet_mos, huang2022voicemos, saeki2022utmos}, and contemporary work further explores multi-dimensional perceptual modeling and broader aesthetic or enjoyment-oriented evaluation~\cite{shi2024samosneuralmosprediction, lian2025apgmosauditoryperceptionguidedmos, nishikawa2025multisamplingfrequencynaturalnessmosprediction, tjandra2025metaaudioboxaestheticsunified}. 
In parallel, preference optimization techniques have been adopted to align models with human judgments in speech and audio generation tasks~\cite{speechalign, cao2025scorespreferencesredefiningmos, ijcai2024p502}.
 
However, MOS may not be always the most suitable protocol; when the target attribute is a domain-specific style construct (e.g., ``anime-likeness'') that does not admit a universally shared absolute scale, evaluators often find absolute scoring inconsistent, while comparative decisions (``which sounds more anime-like?'') yield more reliable signals.
This has led to growing interest in \emph{preference-based} and \emph{pairwise ranking} formulations for subjective evaluation and its automation~\cite{wang2023mospc, ta24_interspeech, shi25b_interspeech}.

\vspace{-2mm}
\section{Data Collecting and Processing for Objective Evaluation of Anime-Like Speech}
\vspace{-1mm}
\subsection{Selection and Preprocessing of Speech Data for Subjective Evaluation} 
\vspace{-1mm}
The speech data for subjective evaluation were constructed from multiple publicly available Japanese corpora that cover anime-related speech as well as general non-anime speech.
Specifically, we used Anim-400k as an anime-derived speech source, and employed ReazonSpeech, consisting mainly of TV programs and everyday speech, as well as the Coco-Nut corpus, which includes diverse speaking styles collected from YouTube, as comparison datasets~\cite{anim400k, reazonspeech, coconut}. All three corpora provide ASR-generated transcriptions.
\begin{table}[t]
\centering
\caption{Statistics of speech samples for subjective evaluation}
\vspace{-2mm}
\label{tab:dataset_stats}
\small
\begin{tabular}{lccc}
\hline
\textbf{Corpus} & \textbf{Train} & \textbf{Test} & \textbf{Total} \\
\hline
Anim-400k~\cite{anim400k} & 1,065 & 250 & 1,315 \\
ReazonSpeech~\cite{reazonspeech} & 828 & 120 & 948 \\
Coco-Nut~\cite{coconut} & 607 & 130 & 737 \\
\hline
Total & 2,500 & 500 & 3,000 \\
\hline
\end{tabular}
\end{table}
\begin{table}[t]
\centering
\caption{Demographics of participants ($n=187$)}
\vspace{-2mm}
\label{tab:demographics}
\small
\begin{tabular}{llr}
\hline
\textbf{Category} & \textbf{Group} & \textbf{Count} \\
\hline
\multirow{4}{*}{Age}
& 20s or younger & 8 \\
& 30s & 48 \\
& 40s & 80 \\
& 50s or older & 51 \\
\hline
\multirow{2}{*}{Gender}
& Male & 142 \\
& Female & 45 \\
\hline
\multirow{3}{*}{Familiarity with Anime}
& Low & 9 \\
& Medium & 103 \\
& High & 75 \\
\hline
\end{tabular}
\vspace{-4mm}
\end{table}
\vspace{-3mm}
\subsubsection{Constructing speech set for evaluation}
\vspace{-1mm}
Since these corpora differ substantially in recording conditions, speaker attributes, and speaking styles, directly using them for evaluation may introduce strong biases. To mitigate these effects, we constructed a speech set through the following three-stage processing pipeline.
\vspace{-1mm}

\myparagraph{Reducing Linguistic Bias Based on Transcriptions} 
Evaluators may unconsciously judge anime-likeness based on textual content rather than acoustics.
To suppress such linguistic bias at the data-selection stage, we performed a text-based screening using a \texttt{Qwen3-30B-Instruct} language model in a zero-shot setting.
For each transcription, the model was prompted to rate how likely the text is to originate from anime subtitles or scripts on a five-point scale (1: very unlikely, 5: very likely), considering writing style, lexical choice, dialogue structure, and emotional exaggeration.
We then retained only utterances with scores $\leq 2$ for the evaluation pool, thus reducing the chance that evaluators can infer anime-likeness from lexical cues alone.
\vspace{-1mm}

\myparagraph{Audio Filtering Based on Quality and Content} 
All audio samples were further filtered for quality and suitability for human listening tests.
After preprocessing with a speech enhancement model Sidon~\cite{nakata2026sidonfastrobustopensource}, we applied three criteria:
(i) transcription quality measured by character error rate between the original ASR transcript and a \texttt{whisper-large-v3} re-transcription,
(ii) duration filtering to keep utterances between \SI{2}{s} and \SI{10}{s}, and
(iii) exclusion of low-quality samples using an existing UTMOS~\cite{saeki2022utmos}, keeping only samples with predicted MOS $> 3$.
This filtering removes samples that are either difficult to recognize, too short/long, or acoustically degraded.
\vspace{-1mm}

\myparagraph{Speaker-Condition-Based Matching}
Finally, to suppress biases related to speaker identity and recording conditions, we extracted speaker embeddings for all samples using an ECAPA-TDNN~\cite{ecapatdnn20} and performed distribution-aware sampling.
Concretely, we first visualized the embedding space via t-SNE to reveal dense regions corresponding to highly similar speaker characteristics.
We then clustered the embeddings and filtered redundantly overlapping regions, so that dominant speaker clusters did not occupy a disproportionate fraction of the set, encouraging speaker diversity before selecting the target utterances.

After the above filtering and matching procedures, we obtained a final pool of 3{,}000 utterances and split them into utterance- and speaker-disjoint train (2{,}500) and test (500) subsets, ensuring that no utterance or speaker appears in both splits. Table~\ref{tab:dataset_stats} shows the corpus-wise distribution for each subset.
\vspace{-3mm}
\subsubsection{Constructing A/B comparison pairs}
\vspace{-1mm}
Consequently, to collect reliable A/B judgments while controlling confounds, we constructed a sparse but balanced set of comparison pairs using both \emph{text similarity} and \emph{speaker similarity}, rather than exhaustively enumerating all possible pairs. 

For each utterance, we retrieve a shortlist of candidate partners by filtering with (i) transcript-embedding cosine similarity computed from transcriptions using a Sentence-Transformers model and (ii) ECAPA-TDNN speaker-embedding cosine similarity, excluding candidates lexically far apart or likely to be the same speaker.
Among the remaining candidates, we rank potential partners by a weighted sum of the two cosine similarities and then greedily add non-duplicate pairs until reaching the target number of comparisons.
To emphasize cross-corpus contrasts, we prioritize cross-source candidates, filling the pair budget with cross-source matches first and allowing same-source pairs only to complete the remaining quota.
In this way, we constructed 12{,}500 pairs for the train split (entirely cross-corpus) and 2{,}500 pairs for the test split (including within-corpus), enabling evaluation under out-of-distribution conditions.

\begin{figure}[t]
    \centering
    \includegraphics[width=0.9\linewidth]{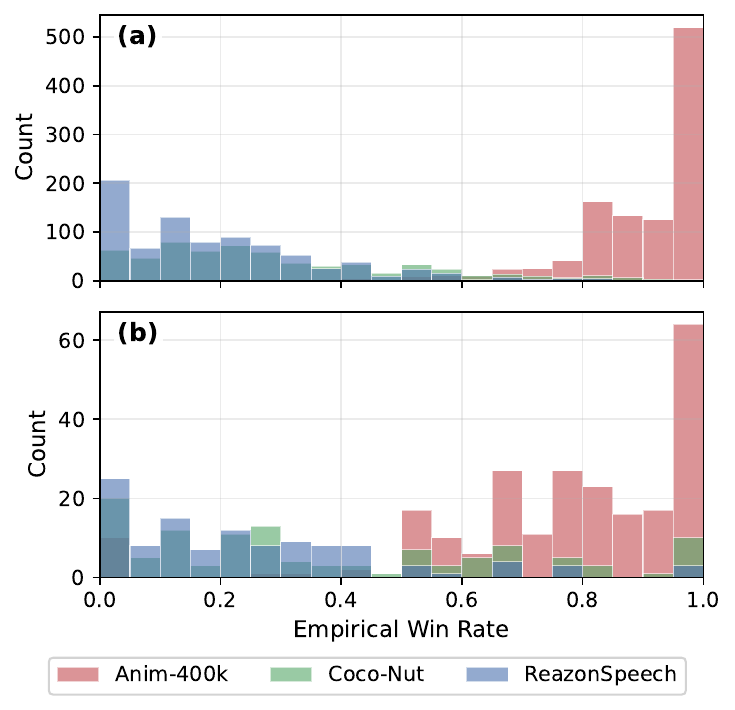}
    \caption{Corpus-wise distribution of empirical win rate for (a) train and (b) test data.}
    \vspace{-5mm}
    \label{fig:hist_ranknet}
\end{figure}

\begin{figure*}[t]
    \centering
    \includegraphics[width=\linewidth]{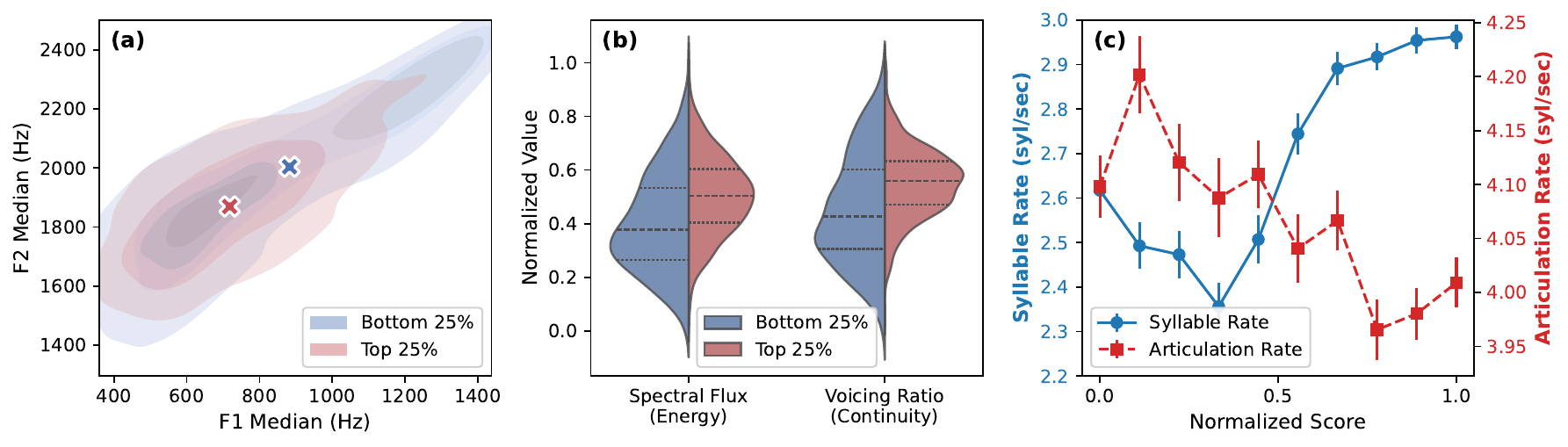}
    \vspace{-4mm} 
    \caption{Visualizations of acoustic proxies for anime-likeness speech. \textbf{(a)} F1-F2 formant space distribution comparing high-win rate (top 25\%) and low-win rate (bottom 25\%) utterances. \textbf{(b)} Split violin plots of normalized spectral flux and voicing ratio distributions for the top and bottom 25\% groups. \textbf{(c)} Overall syllable rate and articulation rate across win rate.}
    \label{fig:perceptual_proxies}
    \vspace{-3mm} 
\end{figure*}

\vspace{-2mm}
\subsection{Subjective Evaluation Design for Labeling}
\vspace{-1mm}
As the primary evaluation protocol, we collected relative A/B preferences from crowd raters recruited remotely via Lancers (paid).
Because anime-likeness is a style-centric impression and may not admit a universally shared absolute scale, we used pairwise comparisons as a practical way to elicit consistent judgments. Each pair received one judgment; we prioritized broad pair coverage (\textasciitilde10 pairs per utterance) over per-pair redundancy, as pairwise preference noise is reduced more efficiently by expanding coverage than by repeated judgments.
\vspace{-1mm}

\myparagraph{Task and instructions.} 
At the beginning of the study, participants read a short instruction page and provided basic demographics (age, gender, and familiarity with anime).
For each trial, they listened to two utterances and selected the one that sounded \emph{more anime-like} according to their own intuition, focusing on voice style rather than content.
To better understand how raters operationalize ``anime-likeness,'' we additionally asked them to write a short free-form description of the cues they personally associate with an anime-like voice.

\vspace{-1mm}
\myparagraph{Session structure and scale.}
One session consisted of 25 A/B trials (approximately 15 minutes).
Each evaluator completed between 1 and 10 sessions (i.e., 25 to 250 trials in total), depending on availability.
In total, we collected 15{,}000 A/B preference labels from 187 evaluators, together with the free-form descriptions.
Table~\ref{tab:demographics} summarizes the distributions of age, gender, and self-reported familiarity with anime.

\begin{table}[t]
\centering
\small
\caption{Distribution of Annotator-Identified Key Features of Anime-Like Speech ($n=187$)}
\begin{tabular}{l r}
\hline
Category & Count \\
\hline
Emotional Explicitness & 62 \\
Timbre Difference & 48 \\
Prosodic Salience & 38 \\
Articulation Clarity & 34 \\
Temporal Control & 5 \\
\hline
\end{tabular}
\label{tab:anime_category_distribution}
\end{table}

\begin{table}[t]
\centering
\caption{Pairwise Concordance Rate (PCR) between acoustic proxies and subjective anime-likeness. $\uparrow$/$\downarrow$: the preferred utterance tends to have a higher/lower value.\textsuperscript{$\dagger$}}
\vspace{-2mm}
\label{tab:pcr_summary}
\small
\begin{tabular}{llr}
\hline
\textbf{Dimension} & \textbf{Proxy Metrics} & \textbf{PCR (\%)} \\
\hline
Emotional Explicitness
& Intensity (Arousal) & 52.2 ($\uparrow$) \\
\hline
\multirow{3}{*}{Timbre Difference}
& $F1$ median (Hz) & 59.6 ($\downarrow$) \\
& $F2$ median (Hz) & 61.5 ($\downarrow$) \\
& $F3$ median (Hz) & 60.1 ($\downarrow$) \\
\hline
\multirow{3}{*}{Prosodic Salience}
& Mean F0 (Hz) & 55.1 ($\downarrow$) \\
& Voicing ratio & 59.5 ($\uparrow$) \\
& Spectral flux (mean) & 60.0 ($\uparrow$) \\
\hline
\multirow{4}{*}{Articulation Clarity}
& Syllable rate & 60.3 ($\uparrow$) \\
& Articulation rate & 53.2 ($\downarrow$) \\
& Pause ratio & 60.0 ($\downarrow$) \\
& Mean pause length & 58.7 ($\downarrow$) \\
\hline
\multicolumn{3}{l}{\textsuperscript{$\dagger$}\scriptsize All bootstrap 95\% CIs ($B{=}1{,}000$) exclude 50\%, confirming significance.}
\end{tabular}
\vspace{-6mm}
\end{table}

\begin{figure*}[t]
    \centering
    \includegraphics[width=0.85\textwidth]{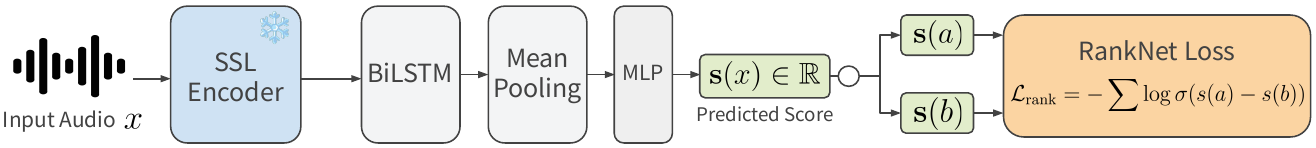} 
    \caption{Architecture of an anime score prediction model}
    \vspace{-4mm}
    \label{fig:animemos_diagram}
\end{figure*}
\vspace{-2.5mm}
\subsection{Experimental Results and Data Analysis}
\vspace{-1.5mm}

\myparagraph{Corpus-wise Distribution of Anime-Likeness Scores}
We first analyze the \textit{empirical win rate} of each utterance to robustly validate corpus-level style differences. As shown in Fig.~\ref{fig:hist_ranknet}, utterances from Anim-400k dominate the high win-rate region, while those from ReazonSpeech and Coco-Nut are concentrated toward the low win-rate region. Anim-400k maintained a 93.2\% win rate against ReazonSpeech and an 88.0\% win rate against Coco-Nut; 
these stark distributional differences confirm that anime-derived speech possesses shared stylistic characteristics that evaluators reliably distinguish from general speech.
\vspace{-1mm}

\myparagraph{Analysis of Annotator-Identified Key Features}
In order to understand the perceptual basis of anime-likeness, we analyzed the qualitative feedback provided by the annotators. 
We pre-defined five dimensions motivated by prior styled-speech studies~\cite{teshigawara2003, starr2015, utsugi2019, ishi2023}: \textit{Emotional Explicitness} (affective intensity), \textit{Timbre Difference} (voice quality), \textit{Prosodic Salience} (pitch/intonation), \textit{Articulation Clarity} (enunciation), and \textit{Temporal Control} (rhythm/pacing).
Free-form responses were collected first to avoid biasing annotators toward predefined options, then assigned to these fixed categories by \texttt{Gemini 3 Pro} as a closed-set classification task (not open-ended generation), which has established LLM reliability~\cite{gilardi2023chatgpt, zheng2023judging}.

Table~\ref{tab:anime_category_distribution} summarizes the distribution of these features, showing that anime-likeness perception is primarily driven by stylistic factors such as emotional explicitness rather than simple acoustic heuristics, with timbre, prosodic salience, and articulatory clarity also contributing.

\vspace{-3mm}
\subsubsection{Relationship between Anime-Likeness and Annotator-Identified Features}
\vspace{-1mm}
To quantify the relationship between annotator-identified dimensions and preference outcomes, we computed the \textit{Pairwise Concordance Rate} (PCR) for each acoustic proxy, with significance assessed via a binomial test against PCR\,=\,0.5. Table~\ref{tab:pcr_summary} summarizes the results.

\vspace{-1mm}
\myparagraph{Emotional Explicitness via Arousal}
We used the arousal value from an Arousal-Valence-Dominance emotion recognition model~\footnote{audeering/wav2vec2-large-robust-12-ft-emotion-msp-dim}.
Arousal yields the weakest concordance (52.2\%), suggesting that perceived ``explicitness'' cannot be reduced to a single affective dimension and likely arises from a combination of prosodic and timbral factors.

\vspace{-1mm}
\myparagraph{Timbre Difference via Formant Statistics}
We analyzed formant frequencies ($F1$--$F3$) using Praat~\footnote{fon.hum.uva.nl/praat/}.
Preferred utterances consistently exhibit \textit{lower} median formants, contradicting the stereotype that anime voices are merely high-pitched.
As shown in Figure~\ref{fig:perceptual_proxies}(a), preferred utterances occupy a lower-frequency resonance space, implying that timbre difference relies on controlled resonance shaping for a fuller vocal quality rather than a simple upward spectral shift.

\vspace{-1mm}
\myparagraph{Prosodic Salience via Pitch and Voicing}
Static pitch height (Mean F0) remains a weak predictor, but preferred utterances exhibit \textit{higher} voicing ratios and spectral flux, indicating a preference for continuous acoustic energy (Figure~\ref{fig:perceptual_proxies}(b)).
Prosodic salience is thus driven by sustained vocal engagement with fewer unvoiced interruptions, not by average pitch.

\vspace{-1mm}
\myparagraph{Articulation Clarity via Rate and Pause}
Preferred utterances have \textit{higher} syllable rates alongside \textit{lower} pause ratios and shorter pauses, while articulation rate (syllables per phonation time) shows negative concordance (Figure~\ref{fig:perceptual_proxies}(c)).
This reveals a paradoxical strategy: continuous rapid flow with minimal pausing, yet deliberate enunciation of individual segments.

\begin{table}[t]
\centering
\caption{Multivariate pairwise prediction via logistic regression on acoustic feature differences (5-fold CV, $N{=}15{,}000$ pairs).}
\vspace{-1mm}
\label{tab:multivariate}
\small
\begin{tabular}{lcrr}
\hline
\textbf{Feature Set} & \textbf{\#Feat.} & \textbf{Acc. (\%)} & \textbf{AUC (\%)} \\
\hline
Emot. Explicit. & 1 & 52.1\small{$\pm$0.6} & 52.9\small{$\pm$0.9} \\
Timbre Diff. & 3 & 60.0\small{$\pm$1.3} & 65.7\small{$\pm$1.4} \\
Prosodic Sal. & 3 & 62.0\small{$\pm$1.2} & 66.0\small{$\pm$1.5} \\
Artic. Clarity & 4 & 59.8\small{$\pm$1.3} & 66.8\small{$\pm$1.4} \\
\hline
All Combined & 11 & 63.4\small{$\pm$1.2} & 69.3\small{$\pm$1.5} \\
\hline
\end{tabular}
\vspace{-4mm}
\end{table}

\vspace{-2mm}
\subsubsection{Multivariate Analysis of Acoustic Features}
\vspace{-1mm}
While the per-feature PCR analysis above reveals individually modest correlations, the question is whether combining all acoustic dimensions yields stronger predictive power. To address this, we trained a logistic regression classifier on the pairwise feature differences for each A/B pair, evaluated via 5-fold stratified cross-validation over all 15{,}000 pairs.

As shown in Table 5, the combined model reaches 69.3\% AUC, confirming an additive benefit of combined dimensions over single dimension. The standardized coefficients reveal that pause ratio and syllable rate are the dominant predictors, with articulation rate showing an opposite sign, corroborating the ``dense flow, deliberate enunciation'' strategy identified in the univariate analysis. Nevertheless, the 69.3\% AUC ceiling suggests that handcrafted acoustic features, even when jointly optimized, fail to capture the full complexity of anime-likeness perception. This motivates the use of learned representations.

\vspace{-2mm}
\section{Designing Score Prediction Model for Anime-Like Speech}
\vspace{-1mm}
As this baseline confirms the multidimensional nature of the target attribute, we further train end-to-end prediction models on top of frozen SSL encoders and show that they substantially surpass the acoustic-feature baseline.

\vspace{-2.5mm}
\subsection{Training Setup for Score Prediction Models}
\vspace{-1.5mm}
As illustrated in Fig.~\ref{fig:animemos_diagram}, the model feeds an input waveform $x$ through a frozen SSL encoder to obtain frame-level features $\mathbf{H}$, which are processed by a BiLSTM and mean-pooled into a fixed-length representation, and finally mapped to a scalar score $s(x) \in \mathbb{R}$ via an MLP:
\[
\mathbf{Z} = \mathrm{BiLSTM}(\mathbf{H}), \quad
\bar{\mathbf{z}} = \frac{1}{T'} \sum_{t=1}^{T'} \mathbf{z}_t, \quad
s(x) = \mathrm{MLP}(\bar{\mathbf{z}}).
\]
Following MOSPC~\cite{wang2023mospc}, the shared network predicts scores $s_a, s_b$ for each pair $(a,b)$, and is trained by minimizing the pairwise logistic loss $-\log\sigma(s_a - s_b)$ against the ground-truth A/B outcome. At inference, the model outputs a scalar score $\hat{s}(x)$, and pairwise accuracy is computed from the sign of $\hat{s}_a - \hat{s}_b$.

\begin{table}[t]
\centering
\small
\caption{Ablation study on SSL backbones for anime-likeness preference prediction (pairwise test set, $N=2500$).}
\label{tab:ranknet_results}
\begin{tabular}{lccc}
\hline
\textbf{Backbone} & \textbf{NLL} & \textbf{Acc.(\%)} & \textbf{AUC(\%)} \\
\hline
wav2vec2~\cite{wav2vec2}   & 0.5139 & 0.7430 & 0.8247 \\
WavLM~\cite{chen2022wavlm}         & 0.4284 & 0.8105 & 0.8944 \\
HuBERT~\cite{hsu2021hubert}       & 0.3852 & 0.8243 & 0.9082 \\
data2vec~\cite{baevski2022data2vec}   & 0.4686 & 0.7709 & 0.8580 \\
\hline
\end{tabular}
\vspace{-4mm}
\end{table}

\vspace{-2mm}
\subsection{Experimental Results and Data Analysis}
\vspace{-1mm}
We evaluate on held-out A/B comparisons, reporting pairwise accuracy, negative log-likelihood (NLL), and ROC-AUC. Table~\ref{tab:ranknet_results} compares four SSL backbones under the same setup. All SSL models substantially outperform the logistic-regression baseline on handcrafted features (69.3\% AUC; Table~\ref{tab:multivariate}), with HuBERT reaching 90.8\% AUC, quantifying the advantage of learned representations over hand-designed proxies. On the within-corpus test pairs (out-of-distribution, unseen in training), the model retains comparable AUC, indicating it captures genuine style rather than corpus identity. 

Among the backbones, masked-prediction models (HuBERT, WavLM) consistently outperform the contrastive model (wav2vec~2.0). Prior analyses have shown that masked prediction encourages encoding of prosodic, paralinguistic, and speaker properties beyond phonetic content~\cite{superb, superb_large_eval}, aligning well with the perceptual dimensions of anime-likeness identified in Section~3.3. data2vec falls between the two groups, consistent with its general-purpose design.

\vspace{-2mm}
\section{Conclusion}
\vspace{-1mm}
We presented AnimeScore, a preference-based framework for evaluating anime-like speech via pairwise data collection, acoustic analysis, and automatic style prediction.
Our analysis revealed that perceived anime-likeness is driven by controlled resonance shaping, prosodic continuity, and deliberate articulation rather than simple heuristics such as high pitch. While handcrafted features reach a 69.3\% AUC ceiling, SSL-based ranking models with masked-prediction objectives substantially surpass it (90.8\% AUC), demonstrating the benefit of learned representations for these multidimensional cues.

The study is limited by moderate data scale, demographic imbalance (76\% male, skewed to 30s--50s), and the absence of model architecture ablations beyond the SSL backbone. Future work will address these while exploring the predictor as a reward signal for generative speech model optimization.

\vspace{-2mm}
\section{Use of Generative AI Disclosure}
\vspace{-1mm}
Generative AI tools were used as part of the described methodology: a large language model (\texttt{Qwen3-30B-Instruct}) for text-based screening of transcriptions (Section~3.1.1) and a large language model (\texttt{Gemini 3 Pro}) for closed-set categorization of free-form annotator responses (Section~3.3). Generative AI was also used for grammar checking and wording refinement of the manuscript. All research ideas, experimental design, analyses, and conclusions are the authors' own.

\bibliographystyle{IEEEtran}
\bibliography{mybib}
\end{document}